\documentclass{ws-ijmpa}

\usepackage{graphicx}

\newcommand{\imag}{\mbox{i}\ }

\begin{document}

\markboth{J. W. Negele}
{Hadron Structure from Lattice QCD}

%
\catchline{}{}{}{}{}
%

\title{Hadron Structure from Lattice QCD }

\author{J. W. Negele$^{1}$}

\author{\begin{flushleft} \quad \quad with LHPC collaborators
 B. Bistrovic$^1$, R. G.
  Edwards$^2$,  G. Fleming$^3$, Ph. H\"{a}gler$^4$,  \\ \quad \quad K. Orginos$^1$, A. Pochinsky$^1$, D. B. Renner$^5$, D. G.
  Richards$^{2}$, and W. Schroers$^6$ \end{flushleft} }

\address{\begin{flushleft}$^1$ Center for Theoretical Physics, Massachusetts Institute
  of Technology, Cambridge, MA 02139, \\ \quad USA\\ 
  $^2$ Jefferson Laboratory MS 12H2, 12000 Jefferson Avenue,
  Newport News, VA 23606,  USA \\ 
  $^3$ Sloane Physics Laboratory, Yale University, New Haven,
  CT 06520, USA\\
  $^4$ Dept.\ of Physics and Astronomy, Vrije Universiteit, De
  Beolelaan 1081,  NL-1081 HV \\ \quad
  Amsterdam, The Netherlands\\
  $^5$ Department of Physics, University of
Arizona, 1118 E.\ 4th.\ Street, Tucson, AZ 85721, USA\\
$^6$ NIC/DESY Zeuthen, Platanenallee 6, D-15738 Zeuthen,
  Germany \end{flushleft}}


\maketitle

\begin{abstract}
The structure of neutrons, protons, and other strongly interacting particles is now being calculated in full, unquenched lattice QCD with quark masses entering the chiral regime.
This talk describes selected examples, including  the nucleon axial charge, structure functions, electromagnetic form factors, the origin of the nucleon spin, the transverse structure of the nucleon, and the nucleon to Delta transition form factor.

\end{abstract}

\section{Introduction}

The challenge in using lattice field theory to solve nonperturbative QCD and understand hadron structure from first principles has been to solve full (unquenched) QCD in the chiral regime of sufficiently light quark masses that one includes the dynamics of the pion cloud and can reliably extrapolate to the physical quark mass. 
%

During the past year, our collaboration has entered the chiral regime by undertaking a series of full QCD calculations\cite{Renner:2004ck} combining computationally economical staggered sea quark configurations generated using the Asqtad improved action by the MILC collaboration\cite{Orginos:1999cr,Orginos:1998ue}  with domain wall valence quarks that have chiral symmetry on a discrete lattice. 
Use of these configurations has enabled us to treat pion masses as light as 359 MeV in volumes with spatial dimension as large as 3.5 fm.  Initial results  are presented below for five pion masses, 359, 498, 605, 696, and 775 MeV, as well as comparisons from the heavy pion world with SESAM full QCD configurations\cite{Eicker:1998sy} using Wilson quarks at pion masses of  744, 831, and 897 MeV.

These proceedings will briefly summarize the experimental observables that are calculable on the lattice and describe selected recent results. More details may be found in recent reviews\cite{Negele:2005,Richards:2005,Negele:2004iu},  recent publications of our group\cite{Dolgov:2002zm,Hagler:2003jd,:2003is,Detmold:2001jb,Renner:2004ck,Bistrovic:2005} and of the QCDSF collaboration\cite{Gockeler:2003jf,Gockeler:2004vx,Gockeler:2005aw,Khan:2004vw}.  

\section{Nucleon Structure}

Since asymptotic freedom renders QCD corrections to high energy scattering small and calculable, high energy lepton scattering provides precise measurements of matrix elements of the light-cone operator \\[.1cm]
$\null \quad \quad \quad {\cal O}(x) \!=\!\int \!\frac{d \lambda}{4 \pi} e^{i \lambda x} \bar
  \psi (\frac{-\lambda  n}{2})\!\!
  \not n {\cal P} e\!^{-ig \int_{-\lambda / 2}^{\lambda / 2} d \alpha \, n
    \cdot A(\alpha n)}\!
  \psi(\frac{\lambda n}{2}),
  $ \\[.1cm]
   where $n$ is a unit vector along the light-cone.   
   Expanding  ${\cal O}(x) $ in local operators via the operator
product expansion generates the tower of twist-two
operators\\[.1cm]
$ \null \quad \quad \quad  {\cal O}_q^{\lbrace\mu_1\mu_2\dots\mu_n\rbrace} = 
{\bar \psi}_q \gamma^{\lbrace\mu_1} \imag{D}^{\mu_2} \dots
  \imag{D}^{\mu_n\rbrace} \psi_q, $\\[.1cm]
whose matrix elements can be calculated in lattice QCD.

The familiar quark distribution $q(x)$ specifying
the probability of finding a quark carrying a fraction $x$ of the
nucleon's momentum in the light-cone frame and the $(n-1)^{th}$ moment of 
this distribution are specified by the
diagonal nucleon matrix elements:\\[.1cm]
$ \null \quad \quad \quad 
\langle P |{\cal O}(x) | P \rangle = q(x),  \quad \quad \quad
\langle P | {\cal O}_q^{\lbrace\mu_1\mu_2\dots\mu_n\rbrace} | P \rangle
\propto \int
dx\, x^{n-1} q(x).
$\\[.1cm]
 Analogous expressions in which the 
twist-two operators contain an additional $\gamma_5$ measure moments of the
longitudinal spin density, $\Delta q(x)$.
 The generalized parton distributions  $ H(x, \xi, t)$ and  $ E(x, \xi, t)$  \cite{Muller:1994fv,Ji:1997ek,Radyushkin:1997ki,Diehl:2003ny} are
measured by off-diagonal matrix elements of the light-cone operator\\[.1cm]
$ \null \quad \quad \quad 
	\langle P' |{\cal O}(x) | P \rangle\! =  \!\langle\!\langle \not n \rangle\!\rangle
	H(x, \xi, t) \nonumber  +  \frac{i \Delta_\nu} {2 m}  \langle\!\langle \sigma^{\alpha \nu} n_{\alpha} \rangle\!\rangle
	E(x, \xi, t),
	$\\[.1cm]
where
$\Delta^\mu = P'^\mu - P^\mu$, $ t  = \Delta^2$, $\xi = -n \cdot \Delta /2$, and
$\langle \!
\langle \Gamma \rangle \! \rangle = \bar U(P') \Gamma U(P)$ for  the Dirac spinor $U$.
Off-diagonal matrix
elements of the tower of twist-two operators
$ \langle P' | {\cal
  O}_q^{\lbrace\mu_1\mu_2\dots\mu_n\rbrace} | P \rangle$ yield moments of the
generalized parton distributions, which in the special case of $\xi$ =
0, are\\[.1cm]
$ \null \quad \quad \quad 
	 \int dx\, x^{n-1} H(x, 0, t)  =    A_{n, 0}(t) , \quad  \int dx\, x^{n-1} E(x, 0, t) =  B_{n, 0}(t),
	$\\[.1cm]
where  $ A_{n, i}(t)$ and $B_{n, i}(t)$ are referred to
as generalized form factors.  The physical observables considered in this work are special cases of these general expressions.

	\begin{figure}[t]
\begin{center}
 \hspace*{-0.2 cm} \hbox{ 
 \includegraphics[scale=0.25,clip=true,angle=0]{g_A-lhpc.eps}
 \hspace*{0cm} \raisebox{-.2cm}
{\includegraphics[scale=0.25,clip=true,angle=0]{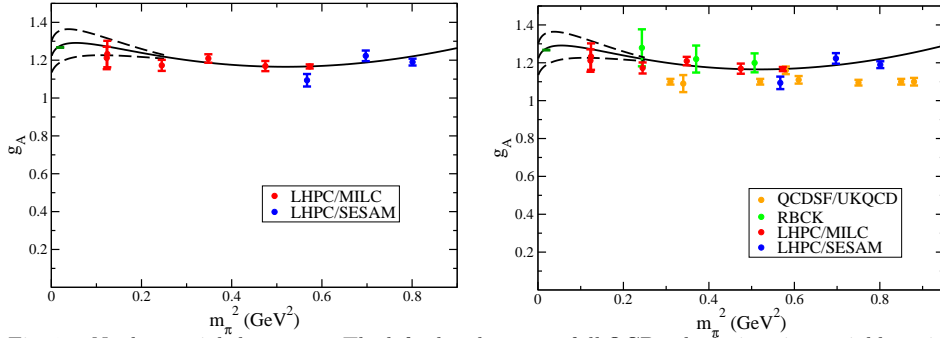}}}
  \vspace*{-0.8cm}
 \caption{Nucleon axial charge, $g_A$. The left plot shows our full QCD calculations in spatial box sizes 1.5 fm (right three error bars), 2.6 fm (next five  points)  and 3.5 fm (second error bar superposed on the left-most point), compared with experiment (data point at far left). The solid line shows a fit to the data using chiral perturbation theory and the dashed lines indicate the error band in this fit.  For comparison, the right graph adds the results by  RBCK in a 1.9 fm box (three lowest new points that touch the solid line) and by QCDSF/UKQCD (remaining new points).}
 \label{fig:gA}
 \end{center}
 \vspace*{-.2cm}
\end{figure}

\medskip
\noindent {\bf Nucleon axial charge}  
\smallskip

 The nucleon axial charge,  $g_A = \langle 1 \rangle_{\Delta q} = \int dx  \Delta q(x)  \propto \int dx \langle \bar q \gamma^\mu \gamma^5 q \rangle $, is a fundamental property of the nucleon governing $\beta$ decay.  It is an ideal test of lattice calculations for several reasons.  The isovector combination $ \langle 1 \rangle_{\Delta q}^{u-d}$ has no contributions from disconnected diagrams, it is accurately measured by neutron $\beta$ decay, and the functional dependence on $m_\pi^2$ is known from chiral perturbation theory\cite{Hemmert:2003cb}. Furthermore, renormalization to account for the difference between regulating quantum field theory with a lattice cutoff and in the continuum can be performed accurately nonperturbatively using the five dimensional conserved current for domain wall fermions. Thus, conceptually, it is a "gold plated" test of our ability to calculate hadron observables from first principles on the lattice, and in addition, since
 it is known to be particularly sensitive to finite lattice volume effects that reduce the contributions of the pion cloud,  it is also a good test for an adequate lattice volume. 
 
   The left panel of Figure~\ref{fig:gA} shows the results of our calculations in the heavy and light pion regimes, together with a fit to a curve defined by chiral perturbation theory\cite{Hemmert:2003cb} that enables us to extrapolate the calculations to the physical pion mass. Note that the extrapolation goes through the experimental point, and that the dashed lines denote the errors in the extrapolation arising from the
errors in the poorly determined counter-term in the chiral expression for $g_A$.
     Also note that at the lowest pion mass, measurements were made in lattice volumes of 2.6 fm and 3.5 fm with statistically indistinguishable results, indicating the absence of finite volume corrections at this lattice size.  For comparison, the right hand graph shows the two other calculations of $g_A$ in full QCD. The three points by the RBCK collaboration\cite{Ohta:2004mg} in a 1.9 fm box are consistent with our results within statistics, whereas the QCDSF/UKQCD results in 1.5-2.2 fm boxes lie uniformly lower than our results and experiment. The axial charge is the most completely analyzed and theoretically controlled of our calculations in the light quark regime, and clearly demonstrates the quantitative potential of lattice QCD with the emerging level of computational resources.

\begin{figure}[t]
\begin{center}
\hspace*{-.3 cm}
 \includegraphics[scale=0.25,clip=true,angle=0]{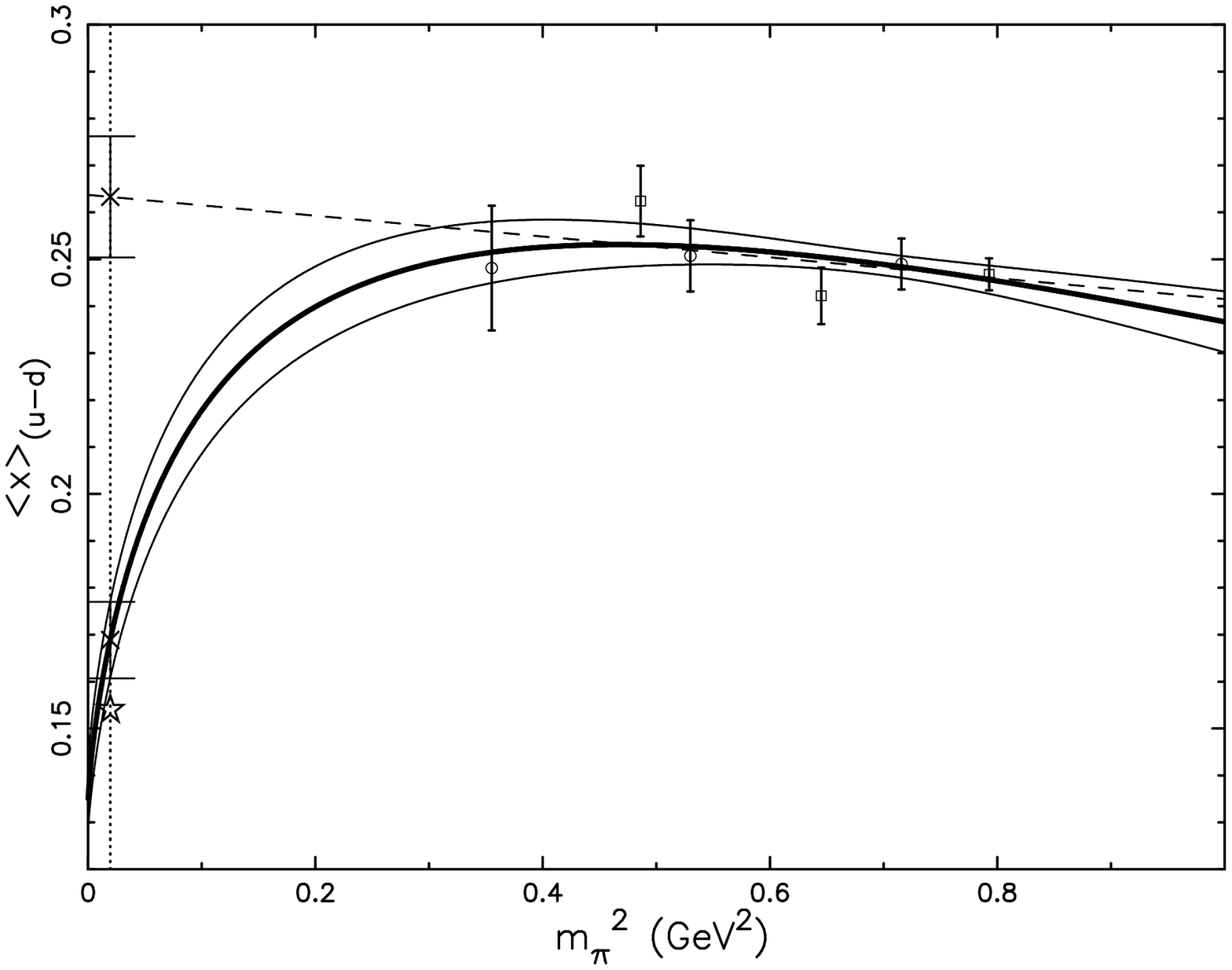}
 \hspace*{.0cm} \raisebox{-0.5cm}{
  \includegraphics[scale=0.31,clip=true,angle=0]{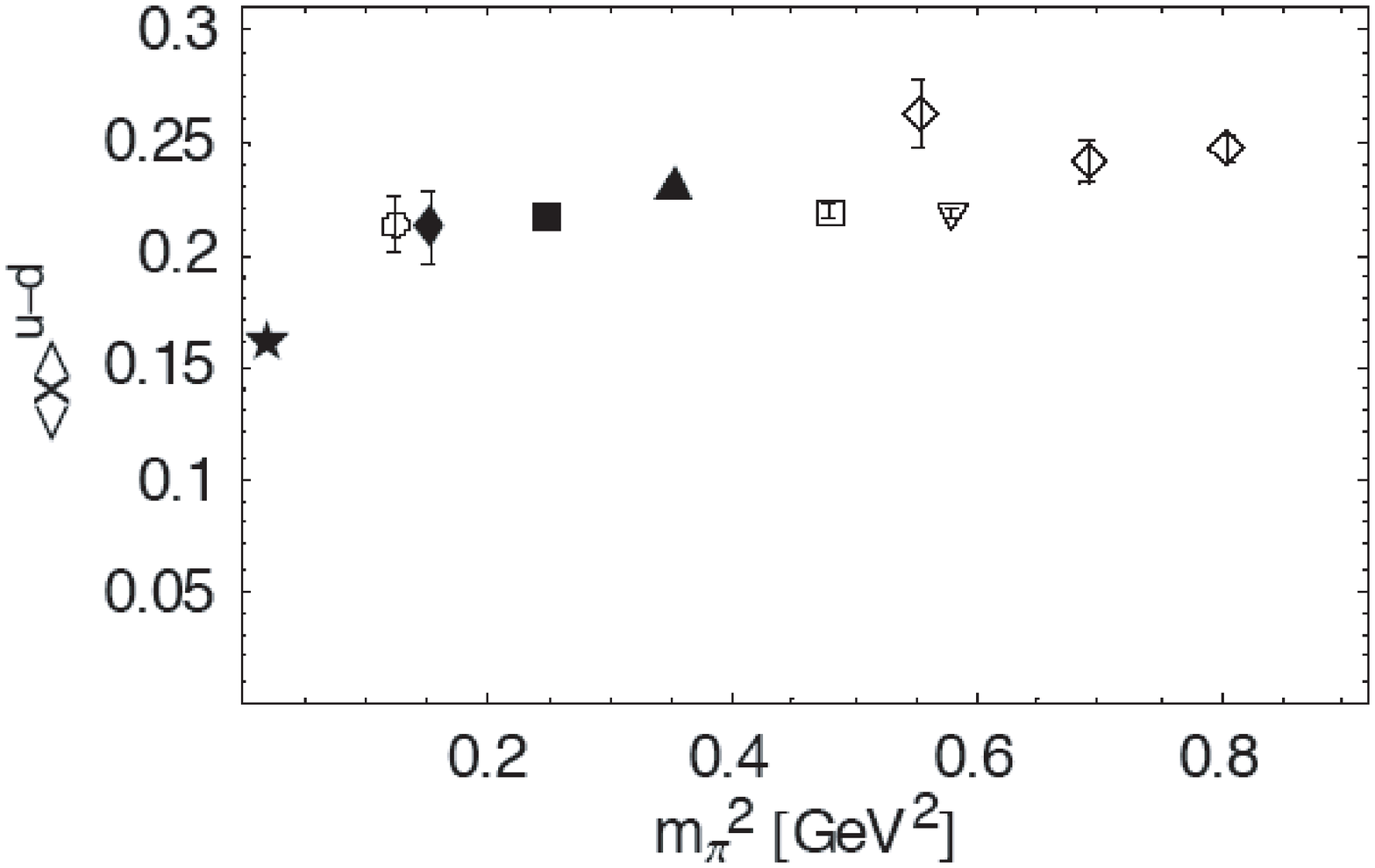} }
 \caption{The left graph shows a curve with chiral perturbation theory behavior at small pion masses adjusted to fit  full QCD and quenched results at large pion masses.  The right plot shows preliminary full QCD calculations extending down to the chiral regime as described in the text. The left-most data point, denoted by an open circle, corresponds to a 359 MeV pion mass and volume 3.5 fm, and the diamond denoting the same mass in a 2.5 fm volume has been shifted slightly to the right for visual clarity.   The star at the left in each graph denotes the experimental value.}
 \label{fig:momfract}
 \end{center}
   \vspace*{-.1cm}
\end{figure}

\medskip
\noindent {\bf Quark momentum fraction}  
\smallskip

The quark momentum fraction, $\langle x \rangle_{q} = \int dx x q(x)  \propto \int dx \bar q \gamma^\mu D^\nu q $,  is  particularly interesting physically, because it reflects the fact that a large fraction of the momentum is carried by gluons rather than quarks.  Figure~\ref{fig:momfract} shows the flavor-nonsinglet difference between the momentum fraction of up and down quarks, which has no contributions from presently uncalculated disconnected diagrams and thus may be compared directly with experiment. 

In contrast to the case of the axial charge, chiral perturbation theory yields a significant dependence of the momentum fraction on the pion mass. The left panel indicates this dependence by extrapolating calculations in the heavy pion regime using  the functional form\cite{Detmold:2001jb} $a [1-\frac{(3g_a^2 + 1 ) m_\pi^2}{(4 \pi f_\pi)^2} ln(\frac{m_\pi^2}{m_\pi^2 + \mu^2})] + b m_\pi^2$. 
%

\begin{figure}[t]
\begin{center}
  \includegraphics[scale=0.35,clip=true,angle=0]{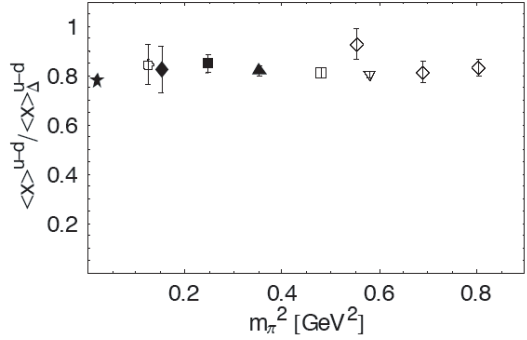} \\[.2cm]
\vspace*{-0.3cm}
 \caption{ QCD predictions for the ratio of the spin-averaged momentum fraction,  
 $\langle x \rangle_q$, to the spin-dependent momentum fraction, $\langle x \rangle_{\Delta q}$, compared with the experimental result (denoted by the star at the left). }
\label{fig:momfractratio}
 \end{center}
   \vspace*{-.4cm}
\end{figure}

The right panel shows the results with chiral valence quarks and MILC configurations  at the five light masses as well as the heavy quark SESAM results denoted by open diamonds.  The light quark calculations are renormalized by calculating in perturbation theory the ratio of the renormalization factor for the operator of interest to the corresponding renormalization factor for the axial current and multiplying by the nonperturbative renormalization constant for the axial current\cite{Bistrovic:2005}.  
%
%
%
Figure~\ref{fig:momfractratio} shows the ratio of the spin averaged momentum fraction, $\langle x \rangle_q$,  to the spin-dependent fraction $\langle x \rangle_{\Delta q}$.  In this case,  the ratio is nearly mass independent and yields excellent agreement with experiment.

\medskip
\noindent {\bf Other observables} 
\smallskip

Space limitations preclude showing other results presented in this talk, so we shall refer briefly here to two other proceedings where these results have recently been published\cite{Negele:2005,Richards:2005}.

The electromagnetic form factors $F_1$ and $F_2$, corresponding to $A_{10}$ and $B_{10}$ defined above, characterize the spatial distribution of charge and current at low momentum transfer and the ability of a single quark to absorb a large momentum transfer and remain in the ground state. Figure 1 of Ref.~\raisebox{-.18cm}{\Large{\cite{Richards:2005}}} shows how the slope of $F_1$ increases as the pion mass decreases, reflecting the increase in the spatial extent of the pion cloud and  how the rms radius extracted from this slope lies on a curve parameterized in terms of chiral perturbation theory that is consistent with experiment at the physical pion mass. Figure 2 of Ref.~\raisebox{-.18cm}{\Large{\cite{Richards:2005}}}
 shows that the $Q^2$ dependence of $F_2$ is consistent with that predicted by the next to leading order light-cone wave
function\cite{Belitsky:2002kj} $F_2 \sim F_1 {\log^2 (Q^2 / \Lambda^2)} / {Q^2}$,
which agrees with recent  Jlab data\cite{Gayou:2001qd}.

The total quark contribution to the nucleon spin\cite{Ji:1996ek} is given by the extrapolation to $t=0$ of  $A^{\mbox{\tiny u+d}}_{20}(t)$ and $B^{\mbox{\tiny  u+d}}_{20}(t)$, and the contribution of connected diagrams is shown in Figure 5 of Ref.~\raisebox{-.18cm}{\Large{\cite{Negele:2005}}} for the case of an 897 MeV pion.  
%
Combined with the calculation\cite{Dolgov:2002zm} of  $\Delta \Sigma = \langle 1 \rangle_{\Delta u} + \langle 1 \rangle_{\Delta d}$, which yields the contribution of the quark spin to the nucleon spin, we learn that considering connected diagram contributions in the heavy quark regime, 68\% of the nucleon spin arises from quark spin, and 0\% arises from quark angular momentum. Similar results have been obtained in Refs. 
 {\cite{Gockeler:2003jf,Mathur:1999uf}}. 
Evaluation of  disconnected diagram contributions and extension to the light quark regime will reveal the full origin of the nucleon spin.

Burkardt has shown\cite{Burkardt:2000za}  that the generalized parton
distribution $H(x, 0, t)$ is the transverse Fourier transform of the quark distribution, $q(x, r_{\perp})$, so the generalized form factor, which can be calculated on the lattice\cite{Negele:2004iu,Hagler:2003jd,:2003is,Gockeler:2004vx}, measures its moments:\\[.1cm]
$\null \quad \quad \quad
A_{n,0}( -\Delta_\perp^2)  
  = \int d^2 r_\perp \int dx\, x^{n-1}  q(x, r_{\perp})\,e^{
 i \vec r_\perp \cdot  \vec \Delta_\perp} .
$\\[.1cm]
Physically, we expect that as $x \to 1$, the transverse distribution approaches a delta-function $\delta(r_\perp)$, and the slope of the form factor therefore approaches zero.
Figure 6 of Ref.~\raisebox{-.18cm}{\Large{\cite{Negele:2005}}}  bears out this expectation, showing that the slope of the form factor $A^{u-d}_{n,0}(t)$ decreases dramatically for higher moments weighting large $x$, and that the transverse rms radius decreases strongly with the average value of $x$.

 The experimental method of choice to reveal the presence of deformation in the low-lying baryons is measuring the N - $\Delta$ transition amplitude, where non-vanishing electric quadrupole (E2) and Coulomb quadrupole (C2) amplitudes are  signatures of deformation in the nucleon,  Delta, or both.  It is convenient to measure the ratio of the electric to magnetic form factors, $R_{EM} = -{\cal G}_{E2}(q^2) / {\cal G}_{M1}(q^2)$ and of the Coulomb to magnetic form factors, $R_{SM} = - |\vec q|{\cal G}_{C2}(q^2) / 2 m_{\Delta} {\cal G}_{M1}(q^2)$.
Figure~4 of Ref.~\raisebox{-.18cm}{\Large{\cite{Richards:2005}}}
 shows the results of a new lattice method that for the first time has the precision to measure non-vanishing $R_{EM}$ and $R_{SM}$ ratios\cite{Alexandrou:2004xn,Alexandrou:2003ea}, and extrapolation of these quenched results in the heavy pion regime to the chiral limit yields results qualitatively similar to experiment.

\medskip
\noindent {\bf Acknowledgements}  
 \smallskip
 
This work was supported by the DOE Office of Nuclear Physics under contracts DE-FC02-94ER40818, DE-FG02-92ER40676, and DE-AC05-84ER40150, the EU Integrated Infrastructure Initiative Hadron Physics (I3HP) under contract RII3-CT-2004-506078, and the DFG under contract FOR 465 (Forschergruppe Gitter-Hadronen-Ph\"anomenologie). Computations were performed on clusters at Jefferson Laboratory and at ORNL using time awarded under the SciDAC initiative. We are indebted to members of the MILC  and SESAM collaborations for the dynamical quark configurations which made our full QCD calculations possible.

\bibliography{SciDAC_Referencesa}

\end{document}